\documentstyle[aps,prb,twocolumn,graphics]{revtex}

\newcommand{\dfn}{\stackrel{\rm def}{=}}

\begin{document}

\title{Symmetry, dimension and the distribution of the conductance
at the mobility edge}
\author{Marc R{\"{u}}hl{\"{a}}nder$^1$, Peter Marko{\v{s}}$^{1,2}$
and C. M. Soukoulis$^1$ \\
$^1$Ames Laboratory and Department of Physics and Astronomy,
Iowa State University, Ames, Iowa 50011\\
$^2$Institute of Physics, Slovak Academy of Sciences,
D{\'{u}}bravsk{\'{a}} cesta 9, 842 28 Bratislava, Slovakia
}
\maketitle

\begin{abstract}
The probability distribution of the conductance at the mobility edge,
$p_c(g)$, in different universality classes and dimensions is investigated
numerically for a variety of random systems. It is shown that $p_c(g)$
is universal for systems of given symmetry, dimensionality, and boundary
conditions. An analytical form of $p_c(g)$ for small values of $g$ is 
discussed and agreement with numerical data is observed. For $g > 1$,
$\ln\, p_c(g)$ is proportional to $(g-1)$ rather than $(g-1)^2$.
\end{abstract}

PACS numbers: 71.30.+h, 71.55.Jv

Disordered systems may show a transition from metallic
states to insulating ones at a mobility edge,\cite{Sou99b,Krm93}
which seperates the two regions.
The probability distribution $p(g)$ of the conductance $g$ at the
critical point of disordered systems undergoing a
localization--delocalization transition is still under 
investigation.\cite{Sha90,Sle97,Wan98b,Ple98,Mks99,Sou99a,Sle00,Bra01,Mtt99}
Previous studies have shown that $p(g)$ depends on the symmetry of the 
system,\cite{Sle97}
its dimensionality,\cite{Mks99} the boundary conditions perpendicular to the 
direction of transport,\cite{Sou99a,Sle00,Bra01} and the amount of 
anisotropy.\cite{Wan98b,Ruh01b} Yet, a complete theory
explaining the form of the distribution is still missing.\cite{Sha90,Mtt99}
Knowing that the conductance distributions are normal and log--normal
in the extended, metallic and the localized, insulating regimes 
respectively, and taking into account the continuous nature of the
Anderson localization--delocalization transition, it seems
reasonable to try to combine the two forms. 

In this letter
we have calculated the probability distribution of the conductance
for a variety of systems of different dimension and symmetry,
give an approximate expression for $p_c(g)$,
compare our numerical results to some analytical approximations,
and present a way of explaining the differences between theory and numerical
results. We find that it 
is necessary to examine the distributions of
the smallest Lyapunov exponents and the relationship between their
respective mean values. We also show that $p_c(g)$
is independent of the particular point chosen on the critical
surface in parameter space, consistent with similar findings on
varying the distribution function of the disorder potential.\cite{Mks94b,Sle01}

We have calculated the conductance distributions at the mobility
edge of a three--dimensional (3D) system with orthogonal symmetry,
a two--dimensional (2D) system with symplectic symmetry, and of 
several 2D systems with unitary symmetry. All these systems possesss 
a mobility edge and are modelled after the
Anderson tight--binding Hamiltonian

\begin{equation}
{\mathcal H} = \sum_{n,\tau} \left| n\tau \right\rangle \varepsilon_n
\left\langle n\tau \right| + \sum _{n,\tau,n',\tau '} \left| n\tau
\right\rangle V_{n,n'} \left\langle n'\tau ' \right|
\end{equation}

\noindent
where $n,n'$ are nearest neighbour sites in the 2D or 3D lattice. 
The variables $\tau,\tau '$ take on values of $1$ or $-1$ for
symplectic systems with spin--orbit interactions, where the hopping integrals
$V_{n,n'}$ thus become $2 \times 2$ matrices; otherwise they are scalars
and the spin ``variables'' have only one value.\cite{vnn}
The site energies $\varepsilon_n$ are always independent of $\tau$.

\begin{figure}[h]
\resizebox{3.0in}{3.0in}{ \includegraphics{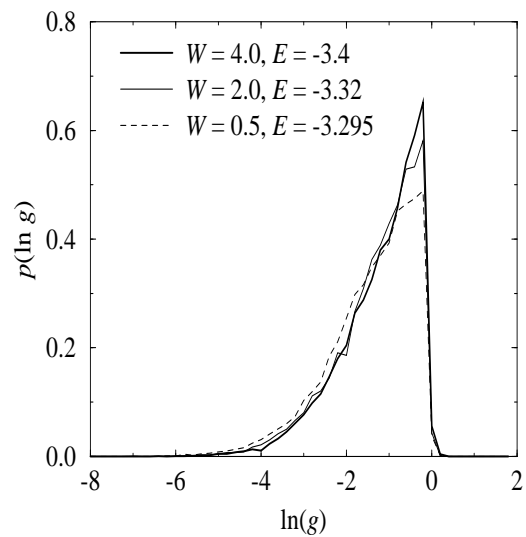} }
\caption{\label{condt1} The conductance distribtuions for three
different critical two--dimensional systems with a magnetic
field perpendicular to the plane.}
\end{figure}

In Fig.\ \ref{condt1} we show three unitary systems with periodic
boundary conditions. A magnetic
field perpendicular to the direction of transport facilitates the 
existence of critical states at the center of each Landau subband.
We investigate the dependence of $p_c(g)$
on the disorder strength $W$.
The flux per unit area $\alpha$ has been kept constant at $\alpha = 1/8$.
For weak disorder, considerable finite
size effects have to be eliminated. 
Even for the system shown with $192 \times 192$ lattice sites (dashed
line in Fig.\ \ref{condt1}) the
distribution still has not completely converged to the form obtained
for the two cases of stronger disorder, where the system size is
only $64 \times 64$ lattice sites (solid lines in Fig.\ \ref{condt1}). 
Anisotropic
systems can be rescaled to the same distribution.\cite{Ruh01b}
Table \ref{stats} contains the averages and standard deviations of
the relevant variables for these ensembles as well as for those we use
in later parts of this paper. 

\begin{figure}[h]
\resizebox{3.0in}{3.0in}{ \includegraphics{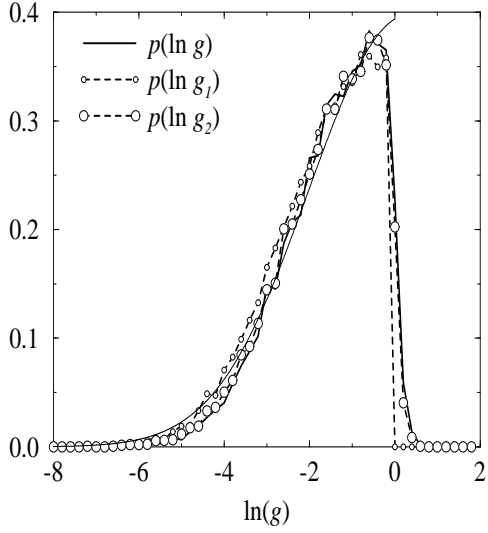} }
\caption{\label{condg1} The distribution of the total conductance $g$ of
cubes with $10 \times 10 \times 10$ lattice sites and periodic
boundary conditions, together with
the distributions for the contributions from the first ($g_1$) and
the first two ($g_2$) channels of the same ensemble. The thin solid
line is the result of Eq.\ \ref{wigner} with $\left\langle z_1 \right\rangle
= 2.825$.}
\end{figure}

We will discuss the transmission properties of a system in terms
of its ``extensive Lyapunov exponents'' $z_i$, where $e^{z_i}$ are
the eigenvalues of ${\mathrm T}^\dagger{\mathrm T}$ and ${\mathrm T}$
is the transfer matrix in the channel representation.
Then, we have for the conductance $g$ (in units of $e^2/h$)\cite{Pic91} 

\begin{equation}
g = \sum_{i=1}^N \frac{1}{\cosh^2 \left( \frac{z_i}{2} \right)}
\end{equation}

\noindent
where $N$ is the number of open channels. The distribution of the
conductance should therefore be discussed in connection with that
of the Lyapunov exponents. The distribution
of the smallest positive Lyapunov exponent $z_1$ can be approximated
by a Wigner distribution with $\beta = 1$,
independently of the actual universality class of the 
system:\cite{Pic91,Mks93}

\begin{equation}
\label{mehta}
p(z_1) \approx \frac{\pi}{2} \, \frac{z_1}{\left\langle z_1 \right\rangle^2}
\, \exp\left(-\frac{\pi}{4} \frac{z_1^2}{\left\langle z_1 
\right\rangle^2}\right) 
\end{equation}

\noindent
where $\left\langle . \right\rangle$ denotes the ensemble average.
This approximation works reasonably well, if $\left\langle z_1 \right\rangle$
is small enough, which is true in two and three dimensions, but not
e.g.\ in four.
Approximating $g \approx g_1 \dfn \cosh^{-2} (z_1/2)$, we can rewrite
this distribution in terms of $\ln (g)$ as

\begin{eqnarray}
p(\ln\,g) & \approx & p(\ln\, g_1)\nonumber\\
 & = & \int_0^\infty \delta \left( \ln (g) 
+ 2\ln\cosh (\frac{z_1}{2}) \right) p(z_1) {\mathrm d} z_1\nonumber\\
\label{wigner}
& \approx & \frac{\pi}{2\left\langle z_1 \right\rangle^2} \,
\frac{z_1}{\tanh \frac{z_1}{2}}\,
\exp\left(-\frac{\pi}{4} \frac{z_1^2}{\left\langle z_1 \right\rangle^2}\right)
\end{eqnarray}

\noindent
evaluated at $\ln(g) = -2\ln\cosh(z_1/2)$.
This obviously neglects contributions to the conductance from 
higher channels and therefore overestimates the distribution
in the range $\ln (g) \leq 0$. Note, that, because $\cosh^2
(z_1/2) \geq 1$ for all $z_1$, $\ln (g_1) \leq 0$. 
One finds that $p(\ln\, g_1)$ is already in reasonable agreement with
$p(\ln\, g)$ indicating that the higher
channels' contributions are small, though not entirely negligible. 
Therefore $p(\ln\, g_1)$ can be
used as a starting point for discussion of the correct 
distribution of the conductance in the range $g \leq 1$. 
Fig.\ \ref{condg1} shows the
numerical results for $10,000$ cubic systems of orthogonal
symmetry with periodic boundary conditions. 
It can be seen clearly from Fig.\ \ref{condg1} that both $p(\ln\, g_1)$ and
Eq.\ \ref{wigner} are in very good agreement with the detailed
numerical results. Also shown is
the distribution $p(\ln\, g_2)$, where $g_2 = g_1 + \cosh^{-2}(z_2/2)$,
which agrees already very well with the distribution of the total
conductance. Squares of symplectic symmetry behave similarly.
Also, systems with hard wall boundary conditions show the
same qualitative behaviour in both 3D orthogonal
systems and 2D symplectic ones.
A summary of the averages and variances of $z_1$ and $z_2$ can be found
in Table \ref{zstats}.

\begin{figure*}[t]
\resizebox{3.0in}{3.0in}{ \includegraphics{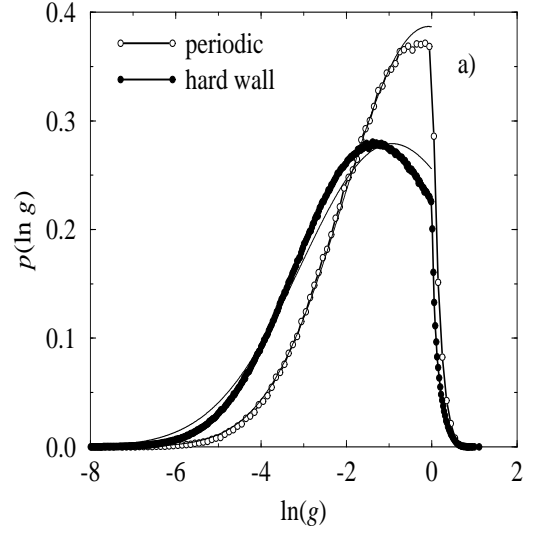} }
\resizebox{3.0in}{3.0in}{ \includegraphics{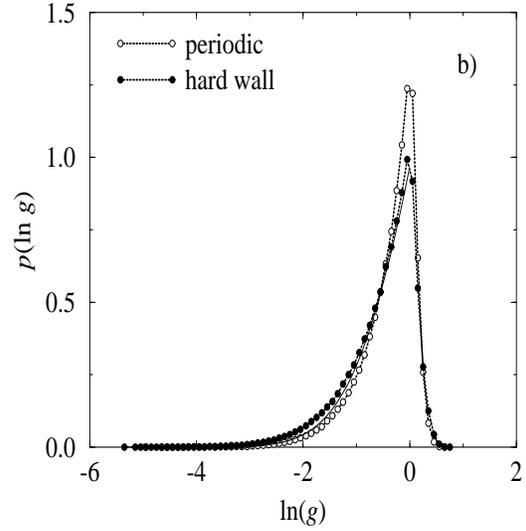} }
\caption{\label{muttalib} The conductance distributions for a)
three--dimensional systems of orthogonal symmetry with $10 \times 10 \times
10$ lattice sites and b) two--dimensional systems of symplectic
symmetry with $40 \times 40$ lattice sites (thick lines). 
The thin lines are fits to the data according to
Eq.\ \ref{wolf} with $\Gamma = \pi/(2.2)^2$, $\Gamma = \pi/(2.55)^2$,
and $\Gamma = \pi/(1.4)^2$ for the 3d  periodic,
3d hard wall, and 2d cases respectively.}
\end{figure*}

Using a different, more elaborate approach, Muttalib and 
W\"{o}lfle\cite{Mtt99} derived for quasi--one--dimensional, weakly
disordered systems a
formula for the critical probability distribution over the
whole range of $g$, including $g > 1$. It can be written as

\begin{equation}
\label{wolf}
p(\ln\, g) = \left\{ \begin{array}{r@{\quad:\quad}l}
\frac{1}{Z}\, \frac{\sqrt{z_1 \sinh\, z_1}}{\tanh\frac{z_1}{2}} \, 
e^{-\frac{\Gamma}{4} z_1^2} & g \leq 1 \\
\frac{\sqrt{2}}{Z} \, g \, e^{-a(g - 1)^2} & g \geq 1
\end{array} \right.
\end{equation}

\noindent
where the formula for the range $g \leq 1$ again needs to be evaluated
at $\ln(g) = -2\ln\cosh(z_1/2)$.
The parameter $\Gamma$ can be used to fit this function to the numerical 
results. ($a$ is a function of $\Gamma$.) 
Taking $\Gamma = \pi/\left\langle z_1 \right\rangle^2$ and noting that
$\sinh(z_1) \approx z_1$ for small $z_1$, the similarity of Eq.\ \ref{wolf}
and Eq. \ref{wigner} is apparent. This suggests replacing $z_1$ in the
prefactor of Eq.\ \ref{mehta} with $\sqrt{z_1\,\sinh(z_1)}$. Preliminary
results show that this actually results in better agreement with data
even for somewhat higher values of $\left\langle z_1 \right\rangle$.
It should be noted though, that in $\Gamma$ instead of the average value 
of $z_1$, one should use the most probable one, which is smaller than the
average value by a factor of about $0.8$ in the case of a Wigner
distribution. Despite their deriving\cite{Mtt99} a distribution
for the whole range of $g$, their formula still overestimates slightly
the weight of the range $g \leq 1$ in 3D systems. However,
in the 2D symplectic case, they slightly underestimate this 
weight. A change in the fitting parameter $\Gamma$ does not remedy this
discrepancy in a satisfactory manner. Fig.\ \ref{muttalib} shows
numerical results together with a fit according to Eq.\ \ref{wolf}.
The first panel shows the distributions for 
3D orthogonal systems with $10 \times 10 \times 10$
lattice sites. In a log--linear plot one can see that Eq.\ \ref{wolf}
increasingly overestimates $p(\ln\, g)$ for $\ln(g) \to -\infty$.
For the 2D symplectic case shown in the second panel,
a fit for $\ln(g)$ close to $0$ results in a very strong underestimation
far from $\ln(g) = 0$. The fit presented for both kinds of boundary
conditions still gives an overall understimation of the portion
of the conductance distribution\cite{note} with $g \leq 1$. 

To understand the qualitatively different behaviours of this
theoretical approach, one has to look at the averages of the
higher Lyapunov exponents.\cite{Mks93} 
In the quasi--one--dimensional, weakly disordered
case for which Eq.\ \ref{wolf} was derived, one has 
$\left\langle z_2 \right\rangle = 2 \cdot \left\langle z_1 
\right\rangle$, independent of dimension, symmetry or boundary
conditions.\cite{Pic91,Mks93,DMP82}
For the 3D orthogonal ensemble, one finds at the critical point
that $\left\langle z_i \right\rangle^2 \propto i$, and thus $\left\langle z_2
\right\rangle$ is significantly smaller than $2 \cdot \left\langle
z_1 \right\rangle$,\cite{Mks93} so that the contribution of the second 
channel is higher than expected from Eq.\ \ref{wolf}, whereas
for the 2D symplectic case, $\left\langle z_2 \right\rangle >
\left\langle z_1 \right\rangle$, so
that the second channel's contribution is smaller than expected.
Compare the values in Table \ref{zstats}, which support these arguments.

\begin{figure}[h]
\resizebox{3.0in}{3.0in}{ \includegraphics{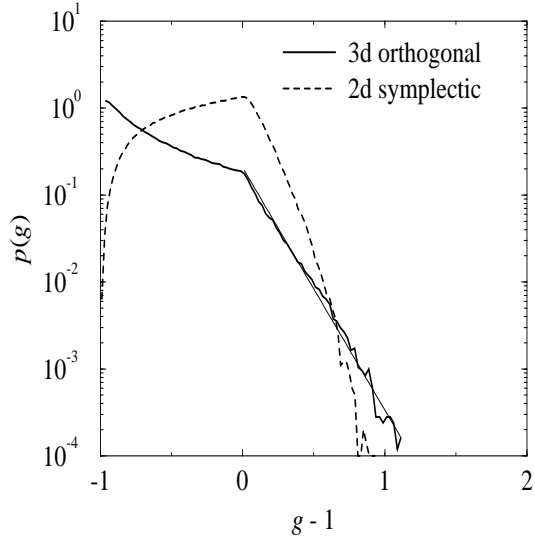} }
\caption{\label{metal} The distribution for an ensemble of $500,000$ cubic
systems (thick line) shows a behaviour $\ln\, p = \mbox{const.} + (g-1)^\alpha$
with $\alpha \approx 1$ (thin line) in the region $g \geq 1$ and a 
discontinuity in its first derivative at $g = 1$. The distribution for
an ensemble of $500,000$ square systems behaves similarly. Both cases
shown use periodic boundary conditions.}
\end{figure}

We also looked at the conductance distribution in the range $g \geq 1$.
In order to have a sizeable ensemble for that range, we took half
a million samples for 2D symplectic systems of $40 \times
40$ lattice sites with both periodic and hard wall boundary conditions as
well as for 3D orthogonal systems of $10 \times 10 \times
10$ lattice sites with periodic boundary conditions. For cubes with
hard wall boundary conditions we even took ten million samples. About 
$20\%$ of the symplectic samples, $6\%$ of the 3D samples
with periodic boundary conditions, and $3\%$ of the 3D
samples with hard wall boundary conditions turn out to have a conductance
bigger than $1$. For the latter ensemble, only about $470$ out of the ten
million samples have a conductance $g > 2$ and only one sample can be
found with $g > 3$. We find that in the range $g > 1$,  $\ln\, p_c(g)$ is at
most linear in $(g-1)$, as can be seen from Fig.\ \ref{metal}.
This is in disagreement with the theory presented
by Muttalib and W\"{o}lfle,\cite{Mtt99} which predicts a quadratic 
dependence with a logarithmic correction, and which therefore expects 
a positive first derivative of $\ln\, p_c(g)$ in $g$.
Finally, Fig.\ \ref{metal} shows that the first derivative of
$p_c(g)$ is discontinuous\cite{cmm} at $g = 1$.
We suppose that this non--analytical behaviour was not taken into account
by the analysis of Muttalib and W{\"{o}}lfle.\cite{Mtt99}

In conclusion, we have shown that the critical distribution of 
the conductance in disordered systems is universal for systems
of given dimensionality, universality class, and boundary conditions.
We show further that for systems of quite different types, the
total conductance is distributed only slightly differently from 
the distribution of the first channel, and give arguments for
the quality of corrections depending on the statistics of the second
channel.  We present a formula for $p_c(\ln\, g)$ which
agrees reasonably well with the numerical results in the range $g \leq 1$.
Finally, we found non--analycity of $p_c(g)$ at $g \approx 1$ and
estimated an exponent of roughly $1$ in $\ln\, p_c(g)$
as a function of $g-1$ rather than the predicted exponent of $2$.

Ames Laboratory is operated for the U.S.\ Department of Energy by
Iowa State University under Contract No.\ W--7405--Eng--82.
This work was supported by the Director for Energy Research,
Office of Basic Science. P.M.\ would like to thank Ames Laboratory
for their hospitality and support and the Slovak
Grant Agency for financial support.

\begin{table}[h]
\caption{\label{stats} The averages and variances of the conductance
and its logarithm for the ensembles we used in this work. O, U, S: orthogonal, 
unitary, symplectic; p, h: periodic, hard wall boundary conditions. 
Unitary systems use periodic boundary conditions.
$N_{\mbox{stat}}$: number of samples.}
\begin{tabular}{|l||r|*{4}{l|}}
System & $N_{\mbox{stat}}$ & $\left\langle g \right\rangle$ & $\sigma_g^2$ & 
$\left\langle \ln(g) \right\rangle$ & $\sigma_{\ln(g)}^2$ \\ \hline\hline
2d U, $W = 4$ & 10,000 & 0.445 & 0.082 & -1.120 & 0.842 \\  \hline
2d U, $W = 2$ & 10,000 & 0.428 & 0.079 & -1.172 & 0.887 \\ \hline
2d U, $W = 0.5$ & 10,000 & 0.393 & 0.078 & -1.306 & 1.027 \\ \hline
2d S, p & 500,000 & 0.749 & 0.088 & -0.401 & 0.283 \\ \hline
2d S, h & 500,000 & 0.691 & 0.108 & -0.531 & 0.418 \\ \hline
3d O, p & 500,000 & 0.391 & 0.108 & -1.418 & 1.282 \\ \hline
3d O, h & 10,000,000 & 0.284 & 0.087 & -1.929 & 1.762 \\
\end{tabular}
\end{table}

\begin{table}[h]
\caption{\label{zstats} The averages and variances of the two smallest
Lyapunov exponents. $N_{\mbox{stat}} = 10,000$.}
\begin{tabular}{|l||*{4}{l|}}
System & $\left\langle z_1 \right\rangle$ & $\sigma_{z_1}^2$ & 
$\left\langle z_2 \right\rangle$ & $\sigma_{z_2}^2$ \\ \hline\hline
2d S, p & 1.424 & 0.621 & 3.987 & 0.924 \\ \hline
2d S, h & 1.635 & 0.811 & 4.065 & 1.186 \\ \hline
3d O, p & 2.825 & 1.918 & 4.965 & 1.829 \\ \hline
3d O, h & 3.411 & 2.475 & 5.518 & 2.132 \\
\end{tabular}
\end{table}


\begin{thebibliography}{99}
\bibitem{vnn}
An external magnetic field enters the Hamiltonian via its vector
potential $\mathbf A$ (${\mathbf \nabla} \times {\mathbf A} = {\mathbf B}$),
which appears in the phases of the hopping integrals:
$V_{n,n'} = t_{n,n'}^0 e^{-2\pi i(e/h)
\int_{{\mathbf r}_n}^{{\mathbf r}_{n'}} {\mathbf A}({\mathbf r})
{\mathrm d}{\mathbf r}}$
The integral connects the lattice sites $n$ (at ${\mathbf r}_n$) and
$n'$ (at ${\mathbf r}_{n'}$) in a straight line. 
The Evangelou--Ziman model\cite{Eva87} incorporates spin--orbit coupling
by using the following hopping integrals:
$V_{n,n'}^{\tau ,\tau '} = t_{n,n'}^0 \left[ \delta_{\tau,\tau '}
+ \mu i \sum_\nu \sigma_{\tau, \tau '}^\nu t_{n,n'}^\nu \right]$
where $\nu = x,y,z$ and $\sigma^\nu$ are the Pauli matrices. The parameter
$\mu$ characterizes the strength of the spin--orbit interaction.
The Ando model\cite{And89} exhibits the same conductance 
distributions.\cite{Mks94b}
The hopping parameter $t_{n,n'}^0$ is chosen constant in all
our systems, where disorder is introduced only by choosing 
the site energies $\varepsilon_n$ independently from a rectangular
distribution of width $W$ centered at $0$. Thus we have in $W$
a measure of the disorder strength. Again, other types of distributions
for $\varepsilon_n$ are known to give the same distributions of
the conductance.\cite{Sle01}
For orthogonal systems we write $V_{n,n'} = t_{n,n'}^0$ for consistency
with the other models.
\bibitem{note}
As mentioned earlier,
the average $\left\langle z_1 \right\rangle$ does not result in
a satisfying fit of Eq.\ \ref{wolf} with $\Gamma = \pi/\left\langle
z_1 \right\rangle^2$, although for 3D systems the
best--fit--value is smaller than $0.8 \cdot \left\langle z_1 \right\rangle$,
whereas for 2D systems it is closer to $\left\langle
z_1 \right\rangle$.
\bibitem{cmm} We found $\partial p_c(g)/\partial g$ of order $-100$ as
$g \to 1^+$ in 3D systems with hard wall boundary
conditions.
\end{thebibliography}
\end{document}